\documentstyle[12pt]{article}
\begin{document}
\large
\begin{centerline}
{\bf DO SUM RULES IN DEEP INELASTIC}
\end{centerline}
\begin{centerline}
{\bf SCATTERING FOLLOW FROM QCD?}
\end{centerline}

 \begin{center} Felix M. Lev \end{center}

\begin{center}
{\it Laboratory of Nuclear Problems, JINR, Dubna, Moscow region 141980,
 Russia (E-mail: lev@nusun.jinr.dubna.su)}
\end{center}

\begin{abstract}
We consider restrictions imposed on the electromagnetic and weak
current operators by Poincare invariance and show that some
assumptions used in deriving the sum rules in deep inelastic
scattering (DIS) have no physical ground. In particular there is
no ground to neglect the contribution of nonperturbative effects
to these operators, even in the Bjorken limit. Therefore the
conclusion that the sum rules in DIS unambiguously follow from QCD is
not substantiated.
\begin{flushleft} PACS: 11.40.Dw, 13.60.Hb \end{flushleft}
\begin{flushleft} Key words: sum rules, deep inelastic scattering,
operator product expansion. \end{flushleft}

\end{abstract}

 {\bf 1.} It is well-known that different sum rules in deep inelastic
scattering
(DIS) are derived with different extent of rigor. For example,
the Gottfried and Ellis-Jaffe sum rules \cite{Got} are essentially
based on model assumptions and in the present paper we will not
consider such sum rules. The sum rules \cite{Adl} were originally derived
in the framework of the current algebra for the time components of the
current operators while the sum rules \cite{Bjor} also involve the
space components. Now the sum rules \cite{Adl,Bjor} are usually considered
in the framework of the operator product expansion (OPE) \cite{Wil}. One
usually thinks that the only important assumption used in deriving these
sum rules is that in the integral over $\omega=1/x_B$ (where $x_B$ is the
Bjorken variable) for the amplitude of the virtual Compton scattering
the contribution of the contour at infinity can be neglected. Since
this assumption is believed to be correct, the sum rules \cite{Adl,Bjor}
have the status
of fundamental relations which in fact unambiguously follow from
QCD. In the present paper we consider some other assumptions in deriving
the sum rules and show that they have no physical ground.

\begin{sloppypar}
 {\bf 2.} If $J(x)$ is the electromagnetic or weak current
operator and $q$ is the momentum transfer then the DIS cross-section
is fully defined by the hadronic tensor
\begin{equation}
W^{\mu\nu}=\frac{1}{4\pi}\int\nolimits e^{\imath qx} \langle P'|
J^{\mu}(x)J^{\nu}(0)|P'\rangle d^4x \quad (\mu,\nu=0,1,2,3)
\label{1}
\end{equation}
where $|P'\rangle$ is the state of the initial nucleon with the
four-momentum $P'$. In the framework of the OPE
\begin{equation}
\int\nolimits e^{\imath qx} J^{\mu}(x)J^{\nu}(0) d^4x = \sum_{i}
C_i(q^2) q_{\mu_1}\cdots q_{\mu_n} O_i^{\mu_1\cdots \mu_n}
\label{2}
\end{equation}
where $C_i(q^2)$ are the $c$-number Wilson coefficients and
$O_i^{\mu_1\cdots \mu_n}$ are some operators.
Then, even if nothing is known about them, it is possible to
determine the $Q^2$ evolution of the structure functions where
$Q=|q^2|^{1/2}$. However for deriving the sum rules one assumes
additionally
that the $O^{\mu_1...\mu_n}_i$ depend only on field operators and their
covariant derivatives at the origin of Minkowski space and have the same
form as in perturbation theory. The basis for twist two operators
contains in particular
\begin{equation}
O_V^{\mu}={\cal N}\{{\bar \psi}(0)\gamma^{\mu}\psi (0)\}, \quad
O_A^{\mu}={\cal N}\{{\bar \psi}(0)\gamma^5\gamma^{\mu}\psi (0)\},
\label{3}
\end{equation}
where ${\cal N}$ stands for the normal product and for simplicity we do
not write flavor operators and color and flavor indices.
\end{sloppypar}

 It is important to note that the OPE has been proved only in perturbation
theory \cite{Br}, its validity beyond that theory is problematic
(see the discussion in ref. \cite{Nov} and references therein)
and perturbation theory does not apply to DIS.
At the same time it is surprising that almost all authors investigating
DIS do not pay attention to the restrictions imposed on the current and
$O$-operators by Poincare invariance. We next consider these restrictions.

 {\bf 3.} Translational invariance of the current operator implies that
 \begin{equation}
J(x)=exp(\imath Px) J(0)exp(-\imath Px)
\label{4}
\end{equation}
where $P$ is the four-momentum operator. In turn, Lorentz invariance
implies that
\begin{equation}
[M^{\mu\nu},J^{\rho}(0)]=-\imath (g^{\mu\rho}J^{\nu}(0)-g^{\nu\rho}
J^{\mu}(0))
\label{5}
\end{equation}
where $M^{\mu\nu}$ are the Lorentz group generators and $g^{\mu\nu}$
is the Minkowski tensor.

 The state $|P'\rangle$ is the eigenstate of the operator $P$ with the
eigenvalue $P'$ and the eigenstate of the spin operators ${\bf S}^2$ and
$S^z$ which are constructed from $M^{\mu\nu}$. In particular,
$P^2|P'\rangle =m^2 |P'\rangle$ where $m$ is the nucleon mass. Therefore
the four-momentum operator necessarily depends on the soft part of the
interaction which is responsible for binding of quarks and gluons in
the nucleon. The Lorentz transformations of the nucleon state are described
by the operators $M^{\mu\nu}$ and therefore in the general case they
also depend on the soft part.

 As noted by Dirac \cite{Dir}, the operators $(P^{\mu},M^{\mu\nu})$ can
be realized in different representations, or, in Dirac's terminology,
in different forms of dynamics. Suppose that the Hamiltonian $P^0$ is
interaction dependent and consider the well-known relation
$[M^{0i},P^k]=-\imath \delta_{ik}P^0$ $(i,k=1,2,3)$. It is obvious
that if all the operators $P^k$ are free then all the operators $M^{0i}$
inevitably contain interaction terms and {\it vice versa},
if all the operators $M^{0i}$ are free then all the operators $P^k$
inevitably contain such terms. According to the Dirac classification
\cite{Dir}, in the instant form the Hamiltonian
$P^0$ and the operators $M^{0i}$  are interaction dependent
and the other six generators of the Poincare group are free, while
in the point form all the components $P^{\mu}$
are interaction dependent and all the operators $M^{\mu\nu}$ are free.
In the front form the operators $P^-$ and $M^{-j}$ ($j=1,2$,
$p^{\pm}=p^0\pm p^z$) are interaction dependent and the other seven
generators are free. The fact that if $P^-$ is the only dynamical
component of $P$ then all the $M^{-j}$
inevitably contain interaction terms follows from the relation
$[M^{-j},P^l]=-\imath \delta_{jl}P^-$. Of course, the physical results
should not depend on the choice of the form of dynamics and in the
general case all ten generators can be interaction dependent.

\begin{sloppypar}
  The usual form of the electromagnetic current operator is
$J^{\mu}(x)={\cal N}\{{\bar \psi}(x)\gamma^{\mu}\psi (x)\}$ and in
particular $J^{\mu}(0)={\cal N}\{{\bar \psi}(0)\gamma^{\mu}\psi (0)\}$.
However such a definition ignores the fact that the product of two
field operators at coinciding points is not a well-defined operator
(strictly speaking, the operator $\psi (0)$ also is not defined
since $\psi (x)$ is the operator-valued distribution). The reader
thinking that it is not reasonable to worry about the mathematical
rigor will be confronted with the following contradiction.
\end{sloppypar}

 The canonical quantization on the hyperplane $x^0=0$ or on the light
cone $x^+=0$ (which leads to the instant and front forms respectively
\cite{Dir}) implies that the operator $\psi (0)$ is free since the
Heisenberg and Schrodinger pictures coincide at $x=0$. Then $J(0)$
is free too and, as follows from Eq. (5), the interaction terms in
$M^{\mu\nu}$ should commute with $J^{\rho}(0)$. If the operators
$M^{\mu\nu}$ are constructed by means of canonical quantization then in
QED the interaction terms and their commutators with $J^{\rho}(0)$
can be readily calculated. The commutators are expressed in
terms of the Schwinger terms \cite{Schw} which cannot be equal to zero
(the corresponding calculation is given in ref. \cite{hep}).
Therefore the conclusion that all the components of $J(0)$ are free is
incorrect.

\begin{sloppypar}
 Moreover, it can be shown \cite{RR} that if the field operators are
quantized, for example, on the hyperplane $x^0=0$ then the operator
${\bf J}(0)$ in QED is necessarily interaction dependent.
Indeed, the generator of the gauge transformations is
$div {\bf E}({\bf x}) - J^0({\bf x})$, and if ${\bf J}(0)$ is gauge
invariant then $[div {\bf E}({\bf x}) - J^0({\bf x}),
{\bf J}(0)]=0$. The commutator $[J^0({\bf x}),{\bf J}(0)]$ cannot be
equal to zero \cite{Schw} and therefore $J^0({\bf x})$ does not
commute with $div {\bf E}({\bf x})$ while the free operator
$J^0({\bf x})$ commutes with $div {\bf E}({\bf x})$.
\end{sloppypar}

 The above examples illustrate the well-known fact that formal
manipulations with local operators in quantum field theory can lead to
incorrect results. For this reason we prefer to rely only upon algebraic
considerations according to which all the components of $J(0)$
cannot be free simply because there is no reason for the interaction
terms in $M^{\mu\nu}$ to commute with the free operators $J^{\rho}(0)$
(see Eq. (5)). Therefore in the instant and front forms
some of the operators $J^{\rho}(0)$ depend on the soft part.
On the other hand, if the operator $J^{\mu}(0)={\cal N}
\{{\bar \psi}(0)\gamma^{\mu}\psi (0)\}$ is free in
the point form, this does not contradict Lorentz invariance but the
operator $J^{\mu}(x)$ in that form necessarily contains the soft part
as follows from Eq. (4).

 The problem of the correct definition of the product of two local
operators at coinciding points is known as
the problem of constructing the composite operators (see e.g. ref.
\cite{Zim}). So far this problem has been solved only in the framework
of perturbation theory for special models. When perturbation theory
does not apply the usual prescriptions are to separate the arguments
of the operators in question and to define the composite operator as
a limit of nonlocal operators when the separation goes to zero (see e.g.
ref. \cite{J} and references therein). Since we do not know how to
deal with quantum field theory beyond perturbation theory, we do not
know what is the correct prescription. Moreover, it is not clear at all
whether it is possible to define local interaction dependent operators
in QCD. Indeed, the dependence of an operator on the soft part implies
that the operator depends on the integrals from the quark and gluon
field operators over the region of large distances where the QCD
running coupling constant $\alpha_s$ is large. It is obvious that such
an operator cannot be local. In particular it is not clear whether
in QCD it is possible to construct local electromagnetic and weak current
operators beyond perturbation theory.

\begin{sloppypar}
 {\bf 4.} Our experience in conventional nuclear and atomic physics tells
that in processes with high momentum transfer the effect of binding is
not important and the current operator can be taken in the impulse
approximation (IA). However this experience is based on the nonrelativistic
quantum mechanics where only the Hamiltonian is interaction dependent
and the other nine generators of the Galilei group are free. In the
usual approaches to DIS it is often claimed that, at least to lowest
order in $1/Q$, the IA is valid in the infinite momentum frame (IMF),
i.e. in the reference frame where $P^{'z}$ is positive and very large.
The IA in the IMF is equivalent to the parton model, and the usual
statement is that the parton model should be modified by taking into
account perturbative QCD corrections. Anyway, one usually thinks that
at least to lowest order in $1/Q$ the soft part of $J(x)$ is
not important as a consequence of asymptotic freedom (i.e. the fact
that $\alpha_s(Q^2)\rightarrow 0$ when $Q^2 \rightarrow \infty$)
and this statement is based on the approach in which the hadronic
tensor (1) is computed by using the OPE in the form of Eq. (2).
\end{sloppypar}

 As noted above, the operator $J(x)$ necessarily depends on the
soft part while Eq. (2) has been proved only in the framework of
perturbation theory. Therefore if we use Eq. (2) in DIS we have to
assume that either nonperturbative effects are not important to some
orders in $1/Q$ and then we can use Eq. (2) only to these
orders (see e.g. ref. \cite{Jaffe}) or it is possible to use Eq. (2)
beyond perturbation theory. The question also arises whether Eq. (2)
is valid in all the forms of dynamics (as it should be if it is the
exact operator equality) or only in some forms.

 In the point form all the components of $P$ depend on the soft part
and hence, as follows from Eq. (4), the integral in Eq. (2)
essentially depends on the soft part. Therefore it is not clear why
the $C_i(q^2)$ remain the same $c$-numbers as in perturbation theory
or, if the dependence on the soft part is moved to the operators
$O_i$ then why they have the same form as in perturbation theory.

 One might think that in the front form the $C_i(q^2)$ will be the
same as in perturbation theory due to the following reasons. The
value of $q^-$ in DIS is very large and therefore only a small
vicinity of the light cone $x^+=0$ contributes to the integral (2).
The only dynamical component of $P$ is $P^-$ which enters into
Eq. (4) only in the combination $P^-x^+$. Therefore the dependence
of $P^-$ on the soft part is of no importance.  These considerations
are not convincing since the integrand is a singular function and the
operator $J(0)$ depends on the soft part in the front form, but
nevertheless we assume that Eq. (2) in the front form is valid. Then,
as follows from Eqs. (1) and (2), the hadronic tensor is defined by
the matrix elements $\langle P'|O_i^{\mu_1\cdots \mu_n}|P'\rangle$.

 Let us consider, for example, the matrix element
$\langle P'|O_V^{\mu}|P'\rangle$. It transforms as a four-vector if the
Lorentz transformations of $O_V^{\mu}$ are described by the operators
$M^{\mu\nu}$ describing the transformations of $|P'\rangle$, or in
other words, by analogy with Eq. (5)
\begin{equation}
[M^{\mu\nu},O_V^{\rho}]=-\imath (g^{\mu\rho}O_V^{\nu}-g^{\nu\rho}
O_V^{\mu})
\label{6}
\end{equation}
It is also clear that Eq. (6) follows from Eqs. (2), (4) and (5).
Since the $M^{-j}$ in the front form depend on the soft part,
we can conclude by analogy with the above consideration that at least
some components $O_V^{\mu}$, and analogously some components
$O_i^{\mu_1\cdot \mu_n}$, also depend on the soft part. Since Eq. (6)
does not depend on $Q$, this conclusion has nothing to do with asymptotic
freedom and is valid even to lowest order in $1/Q$ (in contrast with the
statement of different factorization theorems \cite{ER}). In the
language of Feynman diagrams the fact that one cannot neglect the
soft part of the current operator implies that it is not possible to
separate the diagrams into upper parts describing only the hard
electromagnetic or weak interaction and lower parts depending on
nonperturbative effects, since the struck quark is not free but
interacts nonperturbatively with the rest of the target
(in other words, in the general case "cat ears" diagrams cannot be
reduced to "handbag" ones and even the notion of struck quark is
questionable).

 Since the operators $O_i^{\mu_1...\mu_n}$ depend on the soft part
then by analogy with the considerations in subsection 3 we conclude
that the operators in Eq. (3) are ill-defined and the
correct expressions for them involve integrals from the
field operators over large distances where the QCD coupling constant is
large. Therefore the Taylor expansion at $x=0$
is questionable, and, even if it is valid, the expressions for
$O_i^{\mu_1...\mu_n}$ will depend on higher twist operators which
contribute even to lowest order in $1/Q$. In particular, there is no rule
prescribing that the expression for $O_V^{\mu}$ coincides with
$J^{\mu}(0)$, the expression for $O_A^{\mu}$ coincides with the axial
current operator $J_A^{\mu}(0)$ etc.

 We do not exclude a possibility that (for some reasons) there exist sum
rules which are satisfied with a good accuracy. However the statement
that the sum rules in DIS unambiguously follow from QCD is not
substantiated.

 {\it Acknowledgments.} The author is grateful to G.Salme for the idea
to write this paper
and valuable discussions. The author has also greatly benefited from
the discussions with B.L.G.Bakker, R. Van Dantzig,
A.V.Efremov, B.Z.Kopeliovich, M.P.Locher, M.Marinov,
P.Muelders, N.N.Nikolaev, E.Pace, R.Petronzio, R.Rosenfelder,
O.Yu.Shevchenko, I.L.Solovtsov and H.J.Weber. This work is supported
by grant No. 96-02-16126a from the Russian Foundation for Fundamental
Research.

\end{document}